\documentstyle[aps,preprint]{revtex}

\newcommand{\bq}{\begin{equation}}
\newcommand{\eq}{\end{equation}}
\newcommand{\bqn}{\begin{eqnarray}}
\newcommand{\eqn}{\end{eqnarray}}
\newcommand{\nb}{\nonumber}
\newcommand{\lb}{\label}

\begin{document}
\title{ROTATING CYLINDRICAL SHELL SOURCE FOR LEWIS SPACETIME}
\author{M. F. A. da Silva$^1$\thanks{e-mail: mfas@dft.if.uerj.br}, L.
Herrera$^2$\thanks{e-mail: laherrera@telcel.net.ve}, N. O.
Santos$^{3, 4}$\thanks{e-mail: nos@cbpf.br} and A. Z. Wang$^1$\thanks{e-mail:
wang@dft.if.uerj.br}}
\address{$^1$ Departamento de F\'{\i}sica Te\'orica, Universidade do
Estado do Rio de Janeiro,
CEP: 20550-013, Rio de Janeiro-RJ, Brazil\\
 $^2$ Escuela de F\'{\i}sica, Facultad de Ciencias,
 Universidad Central de Venezuela, Caracas, Venezuela.Postal address:Apartado 80793,Caracas1080A,Venezuela\\
 $^3$ Laborat\'orio Nacional de Computa\c{c}\~ao Cient\'{\i}fica,
CEP: 25651-070, Petr\'opolis-RJ, Brazil \\
 $^4$ Centro Brasileiro de Pesquisas F\'{\i}sicas, CEP: 22290-180,
Rio de Janeiro-RJ, Brazil }
\date{\today}
\maketitle
 
\begin{abstract}
 Rotating thin-shell-like sources for the stationary cylindrically symmetric vacuum solutions (Lewis) are constructed and studied.
It is found, by imposing the non existence of timelike curves in the exterior of the shell, and that the source satisfies the weak, dominant and strong energy conditions
that  the parameters, commonly denoted by $a$ and $\sigma$, are restricted to $0 \le \sigma \le 1/4$ when  $a > 0$, or
$1/4 \le \sigma \le 1/2$  when $a < 0$. 
\end{abstract}

\section{Introduction}

The Lewis metric  represents the most general stationary cylindrically
symmetric solutions to Einstein vacuum equations \cite{Lewis,Kramer}. However,
as it is well known, in the case when all the parameters of the metric are
real, the so called Weyl class, the Cartan scalars corresponding to this
metric are the same as those of the static Levi-Civita spacetime \cite{Levi}
and therefore both metrics, Levi-Civita and Lewis (Weyl class), are locally
indistinguishable \cite{Silva}. This situation is also reflected by the fact
that a coordinate transformation exists \cite{Stachel}, which casts one of the
metrics into the other. Although the consequences implied by this
transformation are physically inadmissible (e.g. periodic time), the
transformation itself is mathematically regular with a non-vanishing Jacobian.

Comments above put in evidence the very peculiar character of the stationarity
 of Lewis metric and the difficulties in the understanding of the physical meaning of its parameters, which has been brought out before \cite{HSP}.

It is our endeavour with this work to delve deeper into this question . With this purpose we shall construct shell-like sources for the Lewis metric.
These shells will be studied and matched to the Lewis spacetime. Doing so, the parameters of the exterior metric will be related to physical properties of the source, and their ranges
of validity somehow restricted by energy conditions. 

It is  worth noticing that although cylindrically symmetric sources are unbounded, and therefore unable of representing real sources of selfgravitating objects, they constitute a
valuable test--bed for numerical relativity, quantum gravity and for probing cosmic censorship and hopp conjecture (see (\cite{Goncalves} and references therein). In this sense, the
kind of source here presented, generalizes the rotating and counter--rotating dust models, usually used in the literature (see (\cite{Jhingham} and references therein). It is also worth
stressing that cylindrically symmetric spacetimes provides an excelent tool for non--perturbative study of departure from spherical symmetry. Finally, let us mention 
that cylindrical symmetry has been used also  to model, as a first order approximation, the rotating matter at the center of galaxies. This 
model, and  the relativistic property of confinement of test particles in the radial direction, might provide another source for the 
extragalactic jet formation \cite{ROpher}.

The paper is organized as follows: in Sec. $II$ we  discuss about the global and local properties of the spacetime, both, 
inside and outside a cylindrical source. Then, we  show that the
solutions with $a > 0,\; \sigma \le \frac{1}{4}$ or  $a < 0,\; \sigma \ge \frac{1}{4}$
are free of CTC's far from the axis of symmetry , where $a$ and $\sigma$ are two of the
four free parameters appearing in the Lewis vacuum solutions, 
with  $a$ and  $\sigma$ being usually related to the angular defect and
the mass per unit length, respectively.  In Sec. $III$, using Taub's method
\cite{Taub}  we construct   cylindrical shell-like sources, by taking
the rotating Minkowski spacetime as the interior of the shell, while in Sec.
$IV$ we solve, for the surface energy momentum tensor obtained in Sec. $III$, the corresponding eigenvalue problem, and then
impose the three energy conditions, weak, strong and dominant. It is found
that these conditions can be satisfied for the solutions with 
$ a > 0,\; 0 \le \sigma \le \frac{1}{4}$ or  $ a < 0,\; \frac{1}{4} \le \sigma
\le \frac{1}{2}$. In Sec. $V$ we calculate   the vorticity of the shell as
well as the energy density per unit length, which   will bring out further the
role of different parameters in the stationarity of the spacetime. The paper is
closed by Sec. $VI$, in which our main conclusions are presented.

\section{The Lewis metric and its local and global properties}

\renewcommand{\theequation}{2.\arabic{equation}}
\setcounter{equation}{0}

The general stationary cylindrically symmetric vacuum solutions of the
Einstein field equations, the Lewis metric, are usually given by
\begin{equation}
\lb{2.1}
ds^2=fdt^2-2kdtd\phi-e^{\mu}(d\varrho^2+dz^2)-ld\phi^2,
\end{equation}
where $f,l,k$ and $\mu$ are functions of $\varrho$ only, being given by
\bqn
\lb{2.2}
f&=& a\varrho^{4\sigma}-\frac{\gamma^2\varrho^{2(1-2\sigma)}}{a},
\;\;\;\; l=\frac{\varrho^2}{f}-\Omega^2f, \nb\\
k&=& -\Omega f, \;\;\;\;\;\;\; e^{\mu}=\varrho^{4\sigma(2\sigma-1)},\;\;\;
\Omega \equiv b+\frac{\gamma\varrho^{2(1-2\sigma)}}{a f},
\eqn
where  $a \;(\not=0),\; b,\; \gamma$ and $\sigma$ are the four free parameters
of the solutions (Note that in this paper we use the notations slightly
different from the ones used in \cite{Lewis,Kramer}). When these parameters
are all real, the corresponding solutions are usually referred to the   Weyl
class, and when they are complex, the corresponding solutions are usually 
referred to the   Lewis class. In this paper we shall restrict ourselves only
to the Weyl class. 

Setting $ b = 0 = \gamma$, the Lewis solutions (with $a > 0$) reduce to the
Levi-Civita solutions \cite{Kramer,Levi}, which represent the gravitational
field produced by a cylindrically symmetric static source,
with $a$ being related to the angular defect and $\sigma$ the mass per unit
length of the cylinder  \cite{Yasuda01},\cite{Yasuda02}. 

In order to give a geometrical meaning to the radial coordinate $\varrho$ we
first transform it into a proper radius $r$ by defining 
\begin{equation} 
\lb{2.4}
\varrho^{2\sigma(2\sigma-1)}d\varrho=dr, 
\end{equation} 
so obtaining
\begin{equation}
\lb{2.5}
\varrho=R^{1/{\Sigma}}, \;\; R={\Sigma} r, \;\; 
{\Sigma} \equiv 4\sigma^2-2\sigma+1.
\end{equation}
With (\ref{2.5}) the metric (\ref{2.1}) becomes
\begin{equation}
\lb{2.5a}
ds^2=Fdt^2-2Kdtd\phi-dr^2-Hdz^2-Ld\phi^2,
\end{equation}
where
\bqn
\lb{2.6}
F&=&aR^{4\sigma/{\Sigma}}-\frac{\gamma^2}{a}R^{2(1-2\sigma)/{\Sigma}}, \;\;\;
H=R^{4\sigma(2\sigma-1)/{\Sigma}}, \nb\\
L&=&\frac{(1-b\gamma)^2}{a}R^{2(1-2\sigma)/{\Sigma}}-ab^2R^{4\sigma/{\Sigma}},
\nb\\
K&=&-\frac{\gamma(1-b\gamma)}{a}R^{2(1-2\sigma)/{\Sigma}}-abR^{4\sigma/{\Sigma}}.
\eqn
Since now  the radial coordinate $r$
defines the proper distance,  without loss of
generality, we shall consider the solutions only in the region $r\in
[0,\infty)$. It can be shown too that the above solutions are singular at
$r=0$ (or equivalently $R=0$) except for the cases $\sigma=0,1/2,\pm\infty$.
In the latter cases, the Riemann tensor vanishes in the region
$r\in(0,\infty)$, and the spacetime is (locally) flat. The singularity at
$r=0$ is usually considered as representing the existence of some kind of
source \cite{Bonnor1}. However, this kind of interpretation is not completely
satisfactory, until some physically acceptable source is found. It is in this
vein that in the following we shall look for shell-like sources for
the above solutions. That is, we shall consider the Lewis solutions valid only
in the region outside of a rotating cylindrically symmetric thin shell, say,
located on the hypersurface $r=r_0$, and then join them to a rotating flat
region in the interior of the shell. By this way, we can consider the Lewis
vacuum field as produced solely by the rotating thin shell. If the matter on
the shell satisfies the weak, strong and dominant energy conditions
\cite{Hawking}, then we shall consider it as physically acceptable source of
the Lewis vacuum solutions. However, we are aware of the fact that a great deal of "exotic" (but still seeming physically meaningful) scenarios
may produce violation of any of the  energy conditions above, and therefore caution should be exerted before ruling out the correspondig sources.

Let us start by noticing that the spacetime inside and outside the shell
must satisfy several physical and geometrical conditions
\cite{Kramer,Yasuda01,Pereira}. In
general, checking those conditions is not trivial. As a matter of fact, only
when the symmetry axis is free of curvature singularities, we know how to
impose them. When it is singular, it is still not clear which
conditions should be required \cite{MacCallum,TW00}. Fortunately,
in the present case since the region inside the shell is assumed to be flat,
the axis is regular. Then, we impose the following conditions:

(i) There must exist an axially symmetric axis, which is usually
characterized by the condition,
\begin{equation}
\lb{2.7}
X\equiv|\xi^{\mu}_{(\phi)}\xi^{\nu}_{(\phi)}g_{\mu\nu}|=\mid g_{\phi\phi}\mid\rightarrow 0,
\end{equation}
as $r\rightarrow 0^{+}$, where we have chosen the radial coordinate such
that the axis is located at $r=0$,  $\phi$ denotes the angular coordinate,
with the hypersurfaces $\phi=0$ and $\phi=2\pi$ being identical,and $\xi^\mu_ {(\phi)}$ is the Killing vector along $d \phi$

(ii) The spacetime near the symmetry axis must be locally flat, which can be
expressed as \cite{Kramer},
\begin{equation}
\lb{2.8}
\frac{X_{,\alpha}X_{,\beta}g^{\alpha\beta}}{4X}\rightarrow 1,
\end{equation}
as $r\rightarrow 0^{+}$ and the comma stands for partial differentiation.
 Note that solutions that fail to satisfy this
condition are sometimes accepted. For example, when the left-hand side of
Eq.(\ref{2.8}) approaches a finite constant, the singularity on the axis can
be related to a cosmic string \cite{VS94}.

These are the conditions that the metric inside of the shell has to satisfy.
For the spacetime outside the shell we impose the conditions:

(iii) {\it No closed timelike curves} (CTC's). In cylindrical spacetimes,
CTC's are rather easily introduced \cite{Hawking}. While the physics of the
CTC's is not yet clear \cite{Hawking1}, we shall not consider this
possibility here and simply require that \cite{Barnes}
\begin{equation}
\lb{2.9}
\xi^\mu_{(\phi)}\xi^\nu_{(\phi)}g_{\mu\nu}<0,
\end{equation}
holds in all the region outside the shell.

(iv) {\it Asymptotical flatness}. Since the source now is a thin shell
located at a finite distance from the axis, we require that the spacetime be
asymptotically flat as $r\rightarrow +\infty$. We note that because of the
cylindrical symmetry, the spacetime can never be asymptotically flat in the
axial direction. Therefore, in the following whenever we mention
asymptotical flatness we mean in radial direction.

(v) {\it No spacetime singularities}. The spacetime outside the shell must
be free of any singularities. By this way, we are sure that the Lewis vacuum
spacetime is indeed produced only by the rotating thin shell. Otherwise, the
singularities may represent additional sources, being a possibility that we
do not consider in this paper.

It can be shown that conditions (iv) and (v) are satisfied by the Lewis
vacuum solutions, while the condition (iii) given by Eq.(\ref{2.9}) becomes
\begin{equation}
\lb{2.10}
L(R)=\frac{(1-b\gamma)^2}{a}R^{4\sigma/{\Sigma}}
[R^{2(1-4\sigma)/{\Sigma}}-R_1^{2(1-4\sigma)/{\Sigma}}]>0,
\end{equation}
where
\begin{equation}
\lb{2.11}
R_1\equiv\left|\frac{ab}{1-b\gamma}\right|^{{\Sigma}/(1-4\sigma)}.
\end{equation}
From the above it is simple to see that Eq.(\ref{2.10}) should hold in the following four cases , far from the axis,   
\bqn
\lb{2.12}
&(a)&\;\;  a>0,\;   \sigma< \frac{1}{4},\;  b\gamma\not= 1,\;
R\in(R_1,\infty);\nb\\ 
&(b)& \;\; a>0,\; \sigma= \frac{1}{4},\;  b\gamma\not= 1,\; 
a^2b^2<(1-b\gamma)^2,\;  R\in(0,\infty);\nb\\ 
&(c)& \;\; a<0, \; \sigma= \frac{1}{4},\;  b \not= 0,\;  a^2b^2>(1-b\gamma)^2, \;
R\in(0,\infty);\nb\\ 
&(d)& \;\;  a<0,\;  \sigma> \frac{1}{4},\; b \not= 0,\;  R\in(R_1,\infty).
\eqn

\section{Matching Lewis spacetime to a cylindrical rotating shell source}

\renewcommand{\theequation}{3.\arabic{equation}}
\setcounter{equation}{0}

In 1937, van Stockum constructed a rotating cylindrically symmetric
dust fluid as the sources   of the Lewis vacuum solutions 
\cite{van37}, and showed that only the solutions (of the Weyl class) with $a >
0$ and $0 \le  \sigma < \frac{1}{4}$ can be produced by such a dust fluid.  

In this paper, we shall consider an infinitely thin cylindrical shell of
anisotropic rotating fluid with a finite radius and we match it to the
exterior Lewis spacetime given by Eqs.(\ref{2.5a}) and (\ref{2.6}). For the
interior of the shell  we assume a rotating Minkowski spacetime, since
it is the only spacetime deprived of energy density. In order to do the
matching we only require the continuity of the metric coefficients across the shell
\cite{Taub}, so allowing us to obtain the most general   rotating thin shell.
Using the same coordinate system as in Eq.(\ref{2.5a}),   the rotating
Minkowski spacetime with angular velocity $\omega$ can be written in the form,
\begin{equation} 
\lb{3.1} 
ds_-^2=(1-\omega^2r^2)dt^2-2\omega
r^2dtd\phi-dr^2-dz^2-r^2d\phi^2. 
\end{equation} 
Indices $-$ and $+$ refer to
interior and exterior spacetimes of the shell, respectively. Clearly the
conditions (\ref{2.7}) and (\ref{2.8}) are satisfied by the metric
(\ref{3.1}). In addition to these two conditions, we also require that the
Killing vector $\xi^{\mu}_{(t)}=\delta^{\mu}_t$ remains timelike in the
interior region of the shell, 
 \begin{equation}
\lb{3.2}
\xi^{\mu}_{(t)}\xi^{\nu}_{(t)}g^{-}_{\mu\nu}= 1 - \omega^2r^2 > 0,
\;\; ( 0 \le r \le r_{0}).
\end{equation}

On the other hand, without loss of generality, we make a
reparametrization of $t$ and $z$,
\begin{equation}
\lb{3.3}
t\rightarrow\frac{t}{A}, \;\; z\rightarrow\frac{z}{B},
\end{equation}
where $A$ and $B$ are constants. Then, the solutions of Eqs.(\ref{2.5a}) and
(\ref{2.6}) become,   
\begin{equation}
\lb{3.4}
ds_+^2=Fdt^2-2Kdtd\phi-dr^2-Hdz^2-Ld\phi^2,
\end{equation}
with
\bqn
\lb{3.5}
F&=&\frac{1}{A^2}\left[aR^{4\sigma/{\Sigma}}-\frac{\gamma^2}{a}
R^{2(1-2\sigma)/\Sigma}\right], \;\;\;
H=\frac{1}{B^2}R^{4\sigma(2\sigma-1)/{\Sigma}}, \nb\\
L&=&\frac{(1-b\gamma)^2}{a}R^{2(1-2\sigma)/{\Sigma}}-ab^2R^{4\sigma/{\Sigma}},
\nb\\
K&=&-\frac{1}{A}\left[abR^{4\sigma/{\Sigma}}+\frac{\gamma(1-b\gamma)}{a}R^{2(1-2
\sigma)/{\Sigma}}\right]. 
\eqn
On the hypersurface $r=r_0$, the first junction condition requires that
\begin{equation}
\lb{3.6}
g^+_{\mu\nu}|_{r_0}=g^-_{\mu\nu}|_{r_0}.
\end{equation}
From the $00$-component of Eq.(\ref{3.6}), we find that 
\begin{equation} 
\lb{3.7}
F(R_0)=\frac{\gamma^2}{aA^2}R^{4\sigma/{\Sigma}}_0[R^{2(1-4\sigma)/{\Sigma}}_2-
R^{2(1-4\sigma)/{\Sigma}}_0] = 1 - \omega^2r^2_{0} > 0,
\end{equation}
where
\begin{equation}
\lb{3.8}
R_0={\Sigma} r_0, \;\;
R_2\equiv\left|\frac{a}{\gamma}\right|^{{\Sigma}/(1-4\sigma)}.
\end{equation}
Eq.(\ref{3.7})   further restricts the validity of the Lewis solutions as
representing the vacuum gravitational field outside a cylindrical source and/or the range of validity of the coordinates of (2.1). 
As a matter of fact, in the cases (a) and (d) given in Eq.(\ref{2.12}), $F(R)$
is always negative when $R$ is sufficiently large. Therefore, in these two
cases the condition $F(R_{0}) >0$  is possible only in certain range of $R$. A
closer investigation shows that the four cases given in Eq.(\ref{2.12}) have
to be further restricted to
\bqn 
\lb{3.9}
&(a)& \;\; a>0, \;\;\;  \sigma< \frac{1}{4},  \;\;\; b\gamma< \frac{1}{2},
\;\;\; R_1<R_0<R_2;\nb\\ 
&(b)&  \;\; a>|\gamma|, \;\;\; \sigma= \frac{1}{4}, \;\;\;
-(a+\gamma)<b<a-\gamma,  \;\;\; b\gamma\neq 1,\;\;\;  R_0>0;\nb\\
&(c)& \;\; -|\gamma|<a<0, \;\;\; \sigma= \frac{1}{4}, 
\;\;\; \gamma-|a|<b<\gamma+|a|,
\;\;\; b\gamma\neq 0, \;\;\; R_0>0; \nb\\
&(d)& \;\; a<0, \;\;\; \sigma> \frac{1}{4}, \;\;\; b\gamma\neq 0, 
\;\;\; b\gamma < \frac{1}{2}, \;\;\; R_1<R_0<R_2.
\eqn
These conditions are sufficient to ensure the absence  of CTC's  outside the source and the timelike
 nature of the Killing vector $\xi^{\mu}_{(t)}$,inside the shell. 

From the above expressions we can see that when
$\sigma \not= 1/4$ the rotating shell can be present only in between the two
cylinders $R = R_{1}$ and $R = R_{2}$ .  It is remarkable to note
that in  the cases (c) and (d) where $a < 0$, the static limit $b = 0 =
\gamma$ is forbidden by the first junction condition. Moreover, in the cases
(a) and (d), there always exists a point $ R = R_{2}$, where 
\bq 
\lb{3.9a} 
F(R) = \cases{ \ge 0, & $ R
\le R_{2}$,\cr < 0, & $ R > R_{2}$.\cr}
\eq
That is, (as expected) the Killing vector $\xi^{\mu}_{(t)} =
\delta^{\mu}_{t}$  changes  from time-like in  the region $R \in [R_{0}
R_{2})$ to space-like in the region $R \in (R_{2}, \infty)$, thereby restricting the range of $R$ (for the whole spacetime) to $(0,R_{2})$.

It should be also noted that the conditions (\ref{3.9}) are valid
not only for the case where a thin shell is  the sole source of the Lewis
metric, but also for the case where the {\em whole} interior region $r \leq
r_0$ is all filled with matter. Then, any kind of matching between a
cylindrical stationary source and the Lewis vacuum spacetime is satisfied by
(\ref{3.9}).

Considering the other components of Eq.(\ref{3.6}),   we
obtain  
\begin{eqnarray}
\lb{3.10}
A&=&\frac{R^{1/\Sigma}_0}{r_0}, \;\;\;
B=R_0^{2\sigma(2\sigma-1)/\Sigma}, \nb\\
a&=&\frac{2(1-b\gamma)^2R^{2(1-2\sigma)/\Sigma}_0}{r_0^2\pm
 \Omega_0}, \nb\\
\omega r_0 &=& -\frac{\gamma
r_0^2}{(1-b\gamma)R^{1/\Sigma}_0}-\frac{2b(1-b\gamma)R^{1/\Sigma}_0}{r^2_0
\pm \Omega_0}.
\eqn
with
\bq
\lb{3.10a}
\Omega_0 \equiv \left[r_0^4+4b^2(1-b\gamma)^2R^{2/\Sigma}_0\right]^{1/2}.
\eq
 
Taub \cite{Taub}  showed that if (\ref{3.6}) is satisfied then the first
derivatives of the metric are in general discontinuous across $r=r_0$, giving
rise to a shell of matter. Following him, we first introduce the quantity
$b_{\mu\nu}$ via the relations
 \begin{equation}
\lb{3.11}
g^+_{\mu\nu,\lambda}|_{r_0}-g^-_{\mu\nu,\lambda}|_{r_0}=n_{\lambda}b_{
\mu\nu},
\end{equation}
where $n_{\lambda}$ is the normal to the hypersurface $r=r_0$, directed
outwards and given by $n_{\lambda}=\delta^r_{\lambda}$. Then, in terms of
$b_{\mu\nu}$,   the energy-momentum tensor (EMT), $T_{\mu\nu}$, of the shell
is given by \cite{Taub},
\begin{equation}
\lb{3.12}
T_{\mu\nu}=\tau_{\mu\nu}\delta(r-r_0),
\end{equation}
where  $\delta(r-r_0)$ denotes the Dirac delta function and $\tau_{\mu\nu}$
the surface EMT, given by
 \begin{equation}
\lb{3.13}
\tau_{\mu\nu}=\frac{1}{16\pi}[b(ng_{\mu\nu}-n_{\mu}n_{\nu})+n_{\lambda}
(n_{\mu}b^{\lambda}_{\nu}+
b^{\lambda}_{\mu})n_{\nu}-nb_{\mu\nu}-n_{\lambda}n_{\delta}b^{\lambda\delta}
g_{\mu\nu}], 
\end{equation}
with $n \equiv n_{\lambda}n^{\lambda}$, and $b \equiv b^{\lambda}_{\lambda}$.
It can be shown that in the present case   the non-vanishing components of
$b_{\mu\nu}$ are 
\begin{equation} 
\lb{3.14}
b_{tt}=2\omega^2 r_0+F^{\prime}_0, \;\; b_{t\phi}=2\omega r_0-K^{\prime}_0,
\;\; b_{zz}=-H_0^{\prime}, \;\; b_{\phi\phi}=2r_0-L^{\prime}_0,
\end{equation}
where a prime stands for differentiation with respect to $r$.
Substituting the above expressions into Eq.(\ref{3.13}), we find that the
surface EMT can be written in the form,
\begin{equation}
\lb{3.15}
\tau_{\mu\nu}=\rho
t_{\mu}t_{\nu}+ q (t_{\mu}\phi_{\nu}+\phi_{\mu}t_{\nu})+p_zz_{\mu}z_
{\nu}+p_{\phi}\phi_{\mu}\phi_{\nu},
\end{equation}
where
\begin{eqnarray}
\lb{3.16}
\rho &=& \frac{1}{16\pi R_0}
\left[1 \mp (1-4\sigma)J(r_{0})\right], \nb\\ 
q &=&  \frac{4\sigma-1}{16\pi R_0}
\left[J^{2}(r_{0})-1\right]^{1/2},\;\;\;
p_z = \frac{(1-2\sigma)\sigma}{4\pi R_0}, \nb\\
p_{\phi}&=&  \frac{1}{16\pi R_0}
\left[1-4\sigma+8\sigma^2 \mp (1-4\sigma)J(r_{0})\right],
\end{eqnarray}
and 
\bqn
\lb{3.16a}
t_{\mu}& =& \delta^t_{\mu}, \;\; z_{\mu}=\delta^z_{\mu}, \;\; \phi_{\mu}=\omega
r_0\delta^t_{\mu}+r_0\delta^{\phi}_{\mu},\nb\\
t_{\mu}t^{\mu} &=& -z_{\mu}z^{\mu}=-\phi_{\mu}\phi^{\mu}=1, \;\;
t_{\mu}z^{\mu}=t_{\mu}\phi^{\mu}=
z_{\mu}\phi^{\mu}=0,
\eqn
with
\bq
\lb{4.7}
J(r_{0}) \equiv \frac{\Omega_{0}}{r^{2}_{0}}.
\eq
The upper sign ``$-$" in Eq.(\ref{3.16}) corresponds to the case $a
> 0$, and the lower sign ``$+$"   corresponds to the case $a
< 0$.

\section{Physical Interpretation of the Surface Energy-Momentum Tensor and
the Energy Conditions}

\renewcommand{\theequation}{4.\arabic{equation}}
\setcounter{equation}{0}

In order to have the physical interpretation for each term appearing in
Eq.(\ref{3.15}), we need to  solve the eigenvalue problem \cite{TW00}, 
\bq
\lb{4.1}
\tau^{\mu}_{\nu} \xi^{\nu} = \lambda \xi^{\mu}.
\eq

Before doing so, we  note that when $\sigma = 1/4$, which
corresponds to the cases (b) and (c) classified in Eq.(\ref{3.9}), we have $q
= 0$ and the surface EMT of Eq.(\ref{3.15})  is already in its canonical form (the same is true when $b=0$).
Then, the three unit vectors $t_{\mu},\; z_{\mu}$ and $\varphi_{\mu}$ are
the corresponding eigenvectors of Eq.(\ref{4.1}).  Thus, now
$\rho$ can be considered as representing the energy density of the matter
shell, and $p_{z}$ and $p_{\varphi}$ the principal pressures along the two
spacelike eigen-directions, defined, respectively,  by  $z_{\mu}$ and
$\varphi_{\mu}$. It can be also shown that the corresponding EMT satisfies all
the three energy conditions \cite{Hawking}. Therefore, it is concluded that
{\em the Lewis vacuum solutions with $\sigma = 1/4$  for both of the two cases
$a > 0$ and $a < 0$,  can be produced by physically acceptable rotating thin
shell}. 

Thus, in the following we need only to consider  the cases (a) and (d) of
Eq.(\ref{3.9}).  In the latter cases,  the system of
equations (\ref{4.1}) will possess nontrivial solutions only when the
determinant ${\rm det}|\tau^{\mu}_{\nu} - \lambda \delta^{\mu}_{\nu}| = 0$,
which   can be written as  \cite{TW00}
\bq 
\lb{4.2}
\lambda (p_{z} - \lambda)\left[\lambda ^{2} - (\rho -
p_{\varphi})\lambda + q^{2} - \rho p_{\varphi}\right] = 0.
\eq 
Clearly, the above equation has four roots,
$
\lambda = 0,\;  p_{z},\; \lambda _{\pm}
$,
where
\bq
\lb{4.4}
\lambda _{\pm} = \frac{1}{2} \left[(\rho - p_{\varphi}) \pm
D^{1/2}\right],\;\;\;
 D \equiv (\rho + p_{\varphi})^{2} - 4 q^{2}.
\eq
It can be shown that the eigenvalue $\lambda = 0$ corresponds to the
eigenvector $\xi_{1} ^{\mu} = n^{\mu}$, where $n^{\mu}$ is the normal
vector to the hypersurface $r = r_{0}$. The
eigenvalue $\lambda = p_{z}$ corresponds to the eigenvector $\xi_{2}^{\mu} =
z^{\mu}$, which represents the pressure of the shell in the $z$-direction.
On the other hand, substituting Eq.(\ref{4.4}) into Eq.({\ref{4.1}), we
find that the corresponding eigenvectors are given, respectively, by
\bq
\lb{4.5}
\xi^{\mu}_{\pm} = (\lambda_{\pm} + p_{\varphi})u^{\mu} + q \varphi^{\mu}.
\eq

In the rest of this section we shall only consider the case $a>0$ in details. For the case $a<0$ we present only the final results since the analysis
is similar to the $a>0$ case.
Thus assuming $a > 0$, we find that 
\bqn
\lb{4.6}
\lambda_{+} + p_{\varphi} &=& \frac{1}{16\pi R_{0}}\left[{\Sigma} 
- (1 - 4\sigma)J(r_{0}) + \sqrt{D}\right],\nb\\
D &=& \frac{1}{(8\pi R_0)^2}
\left[(1-4\sigma)^2+{\Sigma}^2 - 
2{\Sigma}(1-4\sigma)J(r_{0})\right]. 
\eqn
Following \cite{TW00}, we shall further distinguish the three 
subcases: (1) $D > 0$; (2) $D = 0$; and (3) $D < 0$.

\subsection{$D > 0$}

This case splits into two subcases characterized by 
 $\lambda_{+} + p_{\varphi} > 0$ and  $\lambda_{+} + p_{\varphi} < 0$,
respectively.

{\bf Case A.1.1 $\;  \lambda_{+} + p_{\varphi} > 0$}: 

It can be shown that in this subcase, all the three energy conditions are
satisfied for the range of $\sigma$ given by
\bq
\lb{4.15}
0 \le \sigma < \frac{1}{4},
\eq
by properly choosing the radius of the rotating thin shell such that the
following condition is fulfilled,
\bq
\lb{4.15a}
\sqrt{D} \ge \frac{\sigma(1-2\sigma)}{4\pi R_{0}}.
\eq 
On the other hand, it can be also shown that $\lambda_{+} + p_{\varphi} > 0$ is automatically
satisfied, once $D>0$ and (\ref{4.15}) are fulfilled. 
Therefore, it is concluded that {\em all the Lewis vacuum solutions with $a >
0$ and $0 \le \sigma < 1/4$ can be produced by a physically acceptable rotating
shell}.

{\bf Case A.1.2 $\; \lambda_{+} + p_{\varphi} < 0$}: 
It is not difficult to show that in this case none of the three energy
conditions is satisfied. Thus, it is concluded that in the present case there
does not exist physically acceptable  rotating thin shell such that the
conditions $D>0$ and $\lambda_{+} + p_{\varphi} < 0$ are satisfied.

\subsection{$ D = 0$} 

In this case  we find that
when $ a > 0$ and $D = 0$ the weak and strong
energy conditions are satisfied  when $\sigma \in (0, 1/4)$, {\em and
there does not exist any value of $\sigma$ in the range $\sigma \in
(-\infty, 1/4)$, for which the dominant energy condition is satisfied}.

\subsection{$ D < 0$} 

We find that  
{\em in the present case  the weak and strong energy  conditions  are
violated for values of $\sigma$ within the range $ 0 < \sigma < 1/4$}. On the other hand, it can be shown that now the dominant energy
condition requires, 
\bq
\lb{4.32c}
0 \le \sigma < \frac{1}{4},\;\;\;\;\;
\frac{{\Sigma}^{2} + (1 - 4\sigma)^{2}}{2(1-4\sigma){\Sigma}}
< J(r_{0}) \le \frac{8\sigma^{2} - 4\sigma + 1}{1 - 4\sigma}.
\eq 

In review of all the above, it is  concluded that {\em for the case
$a > 0$ the Lewis vacuum solutions  can be produced by physically acceptable
rotating cylindrical thin shells  for   $ 0\le \sigma < 1/4$. Moreover,
to this   range of $\sigma$ the radius of the shell has to be chosen such
that the condition $D > 0$ is satisfied, in which the surface EMT can be
diagonalized and given in its canonical form}.

A similar analysis shows   that {\em for the case
$a < 0$ the Lewis vacuum solutions  can be produced by physically acceptable
rotating cylindrical thin shells for   $ 1/4 < \sigma \le 1/2$, by
properly choosing    the radius of the
shell so that the condition $D > 0$ is satisfied,
for which the surface EMT can be diagonalized and given in its canonical form}.

\section{The vorticity of the shell and its energy per unit length}

\renewcommand{\theequation}{5.\arabic{equation}}
\setcounter{equation}{0}

The four velocity of a comoving observer in the system of the chosen
coordinates  is given by 
\begin{equation}
\lb{5.1}
u^{\mu}=\frac{1}{\sqrt{g_{tt}}}\delta^{\mu}_{t},
\end{equation}
for which it can be shown that the vorticity tensor $\omega_{\alpha\beta}$  
has only two non-vanishing components, given  by \bqn
\lb{5.2}
\omega_{\alpha\beta} &=&
u_{[\alpha;\beta]}+u_{[\alpha;\mu}u^{\mu}u_{\beta]}\nb\\
&=&
\frac{g'_{tt}}{2\sqrt{g_{tt}}}\left(\delta^{t}_{\mu}\delta^{r}_{\nu} -
\delta^{r}_{\mu}\delta^{t}_{\nu}\right)  
+ \frac{g'_{t\varphi}}{2\sqrt{g_{tt}}}
\left(\delta^{r}_{\mu}\delta^{\varphi}_{\nu}  -
\delta^{\varphi}_{\mu}\delta^{r}_{\nu}\right). 
\eqn
Then, the vorticity vector $\omega_{\alpha}$ takes the form 
\begin{equation}
\lb{5.3}
\omega^{\alpha}=\frac{\epsilon^{\alpha\beta\gamma\delta}}{2\sqrt{-g}}
u_{\beta}\omega_{\gamma\delta} = 
\frac{1}{2\sqrt{-g}}\left({g'}_{t\varphi} 
- \frac{{g}_{t\varphi}{g'}_{tt}}{g_{tt}} \right)\delta^{\alpha}_z.
\end{equation} 
Calculating the above quantity  at $r=r_0$ using the external metric
(\ref{3.4}), we find that
\begin{equation}
\lb{5.4a}
\omega^{\alpha}_{+}(r_{0}^{+})= \frac{\gamma
(4\sigma-1)r_0^2}{(1-\omega^2r_0^2)R^{1+1/\Sigma}}\delta^{\alpha}_z,
\end{equation}
while for the interior metric (\ref{3.1}), we have 
\begin{equation}
\lb{5.4b}
\omega^{\alpha}_{-}(r_{0}^{-})=\frac{\omega}{1-\omega^2r_0^2}\delta^{\alpha}_z.
\end{equation}
Clearly, now we have $\omega^{\alpha}_{-} \not= \omega^{\alpha}_{+}$.
The reason   is that the derivatives of the metric are discontinuous across
the shell $r=r_0$.

On the other hand, considering Israel's definition \cite{Israel} of energy
density per unit length $\mu$, from Eqs.(\ref{3.15}) and (\ref{3.16})  we find
that, 
\bqn
\lb{6.1}
\mu &=&
\int^{\infty}_0\int^{2\pi}_0(T^t_t-T^r_r-T^z_z-T^{\phi}_{\phi})\sqrt{-g}
drd\phi\nb\\
&=&
\frac{\sigma}{{\Sigma}}+(1-4\sigma)\frac{b\gamma}{2{\Sigma}}.
\eqn  
The tangential velocity $\omega r_0$ given by (\ref{3.10}) for  the case $a >
0$, up to first order, $O(b)$ and $O(\gamma)$, becomes
\begin{equation}
\lb{6.2}
\omega r_0\approx-\frac{bR_0^{1/{\Sigma}}}{r^2_0}-\frac{\gamma
r_0^2}{R_0^{1/{\Sigma}}}.
\end{equation}
The first and second terms in the right hand side of (\ref{6.2}) correspond to the
tangential velocity of the shell due to $b$, and $\gamma$, respectively.
Then we can see that $b\gamma\sim$(tangential velocity of the shell)$^2$.
Hence, the second term in the right hand side of Eq.(\ref{6.1}), due to
rotation, can be associated to the kinetic energy of the shell.

\section{Discussions and Conclusions}

In this paper, we first studied the local and global properties of 
the stationary cylindrically symmetric general vacuum solutions 
(Lewis), and found that the condition for the non-existence of closed time-like curves outside the shell can be  satisfied if $a > 0,\; \sigma \le 1/4$ or  $a < 0,\; \sigma \ge 1/4$.

To further study the solutions, we also constructed rotating thin-shell-like
sources, by assuming that the spacetime inside the shell is flat. It was shown
that such constructed cylindrical shells  can satisfy the three energy
conditions, weak, dominant and strong, when $a > 0,\; 0 \le \sigma \le
1/4$ or  $a < 0,\; 1/4 \le \sigma \le 1/2$. It was also found that in the
latter cases the corresponding surface EMT can be diagonalized and takes its canonical form. Moreover,  in the cases $a < 0$ the first junction
condition does not allow the static limit $b = 0 = \gamma$ [cf.
Eq.(\ref{3.9})].

The vorticity of the rotating shell and its mass per unit length were also
calculated. When $\gamma = 0$,  the vorticity of the shell calculated
from outside, vanishes as can be seen from Eq.(\ref{5.4a}), while 
the energy per unit length as given by (\ref{6.1}) is the same as that in
the corresponding   static case \cite{Wang}. However the
stationarity of the spacetime manifests itself through the dragging of a
gyroscope at rest in the frame where (\ref{2.1}) takes a diagonal form
\cite{moregyro}.
This situation is reminiscent of the behaviour of a gyroscope in the field
of a charged magnetic dipole. In this latter case, even though the metric
is static, dragging of inertial frames appears and is explained as due to the
presence of a flow of electromagnetic energy in the angular direction
\cite{Bonnor}. In our case also, even if the vorticity of the shell vanishes, there is still a flow of energy (if $b\not=0$ ) along $\phi^{\mu}$,
which might be interpreted as the ''source`` of the dragging.

It is worth mentioning that we have ensured no CTC's and energy conditions being satisfied simultaneously 
in our models by restraining the range of $\sigma$ to $0 \le \sigma \le 1/4$, (if
$a>0$). It remains to be proved if, and to what extent, this range can be safely extended. Since physically reasonable sources for the Levi-Civita spacetime have been found for
$\sigma>1/4$ (see
\cite{Yasuda01}, \cite{Yasuda02} and references therein), it could be conjectured that for sufficiently small values of $b$ and $\gamma$, this is also possible for Lewis. However a
bifurcation might be present , but  we conclude without giving a definite answer to that question.

Finally, we would like to stress once again, that in spite of the highly idealized nature of any cylindrically symmetric source (and the resulting space--time), these have been, and are
currently, considered, as useful tools in the study of many, physically relevant aspects of self--gravitating systems (see \cite{Goncalves} , \cite{Jhingham}, \cite{ROpher} and
references therein).

\section*{Acknowledgements}

The financial assistance from UERJ (MFAdaS, AZW), CNPq (NOS, AZW),  M.C.T.,Spain, BFM2000-1322  and from C\'atedra-FONACIT, under grant
2001001789.(LH) is gratefully acknowledged.


\begin{thebibliography}{99}

\bibitem{Lewis}T. Lewis,   { Proc. Roy. Soc. London}, {\bf 136}, 176 (1932).

\bibitem{Kramer}D. Kramer, H. Stephani, M.A.H. MacCallum,  and E. Herlt,
{Exact Solutions of Einstein's Field Equations,} (Cambridge University
Press, Cambridge, 1980).

\bibitem{Levi} T. Levi-Civita   { Rend. Acc. Lincei}, {\bf 26}, 307 (1917).

\bibitem{Silva} M.F.A. da Silva, L. Herrera, F.M. Paiva, and N.O. Santos,
{ Gen. Relativ. Grav.} {\bf 27}, 859 (1995).

\bibitem{Stachel} J. Stachel, {Phys. Rev. } {\bf D26}, 1281 (1982);E Frehland {Comm. Math. Phys.}, {\bf 23}, 127 (1971);
W B Bonnor, {J. Phys.A} {\bf 13}, 2121 (1980).

\bibitem{HSP} L. Herrera, F.M. Paiva, and N.O. Santos, { Class.
Quantum Grav.} {\bf 17}, 1549 (2000).

\bibitem{Goncalves} S. Goncalves,{gr--qc/0202042}, (2002).

\bibitem{Jhingham} T. Apostolatos and K. S. Thorne, {Phys. Rev.} {\bf 46}, 2435 (1992); S. Goncalves and S. Jhingham {gr--qc/0203077}, (2002); B. Nolan {gr--qc/02044036}, (2002).

\bibitem{ROpher} R. Opher, N O Santos and A Wang, {J. Math. Phys.} 
{\bf 37}, 1982 (1996).

\bibitem{Taub} A.H. Taub, {J. Math. Phys.} {\bf 21}, 1423 (1980).
 
\bibitem{Yasuda01}  L. Herrera, N.O. Santos, A.F.F. Teixeira, and A.Z. Wang, {Class. Quantum Grav.} {\bf 18}, 3847 (2001).

\bibitem {Yasuda02} A.Y. Miguelote,   M.F.A. da Silva, A.Z. Wang, and N.O.
Santos, {Class. Quantum Grav.} {\bf 18}, 4569  (2001).

\bibitem{Bonnor1} W.B. Bonnor, { Gen. Relativ. Grav.} {\bf 24}, 551 (1992);
W.B. Bonnor, J.B. Griffiths, and M.A.H. MacCallum, { Gen. Relativ.
Grav.} {\bf 26}, 687 (1994); M.F.A. da Silva, L. Herrera, F.M. Paiva, and
N.O. Santos, {J. Math. Phys.} {\bf 36}, 3625 (1995); S. Haggag and F. Dsokey,
{ Class. Quantum Grav.} {\bf 13}, 3221 (1996).

\bibitem{Hawking} S.W. Hawking and G.F.R. Ellis, {The Large Scale
Structure of Spacetime}, (Cambridge University Press, Cambridge, 1973) pp.
88-96.

\bibitem{Pereira} P.R.C.T. Pereira, N.O. Santos,  and A.Z. Wang, {
Class. Quantum Grav.} {\bf 13}, 1641 (1996).


\bibitem{MacCallum} M.A.H. MacCallum and N.O. Santos,{ Class.
Quantum Grav.} {\bf 15}, 1627 (1998); M.A.H. MacCallum, {Gen. Relativ. Grav.}, {\bf 30},
131 (1998).

\bibitem{TW00} P.R.C.T. Pereira   and A.Z. Wang, {Gen. Relativ. Grav.}
{\bf 32}, 2189 (2000).

\bibitem{VS94}  A. Vilenkin and E.P.S.  Shellard, {\em Cosmic Strings
and other Topological Defects}, (Cambridge University Press, Cambridge,
1994).

\bibitem{Hawking1} S.W. Hawking, {Phys. Rev. } {\bf D46}, 603 (1992).

\bibitem{Barnes} A Barnes, {Class. Quantum Grav.} {\bf 17}, 2605 (2000).

\bibitem{van37} W.J. van Stockum, Proc. R. Soc. Edin. {\bf 57}, 135 (1937).


\bibitem{Israel} W. Israel, { Phys. Rev. } {\bf D15}, 935 (1977).

\bibitem{Wang} A.Z. Wang, M.F.A. da Silva, and N.O. Santos, {Class.
Quantum Grav.} {\bf 14}, 2417 (1997).


\bibitem{moregyro}L. Herrera and N.O. Santos, {J. Math. Phys.} {\bf 42}, 4956 (2001).

\bibitem{Bonnor} W.B. Bonnor, {Phys. Lett. } {\bf A158}, 23 (1991).
\end{thebibliography}
\end{document}